\documentclass[twocolumn,pra]{revtex4}% Physical Review A

\usepackage{graphicx}% Include figure files
\usepackage{dcolumn}% Align table columns on decimal point
\usepackage{bm}% bold math

\begin{document}

\preprint{APS/123-QED}

\title{The dynamics of quantum phases in a spinor condensate}
\author{S. Yi, \"O. E. M\"ustecapl{\i}o\u{g}lu, and L. You}
\affiliation{School of Physics, Georgia Institute of
Technology, Atlanta, GA 30332-0430, USA}
\date{\today}
\begin{abstract}
We discuss the quantum phases and their diffusion dynamics in a
spinor-1 atomic Bose-Einstein condensate. For ferromagnetic
interactions, we obtain the exact ground state distribution of the
phases associated with the total atom number ($N$), the total
magnetization (${\cal M}$), and the alignment (or hypercharge)
($Y$) of the system. The mean field ground state is stable against
fluctuations of atom numbers in each of the spin components, and
the phases associated with the order parameter for each spin components
diffuse while dynamically recover the two broken continuous symmetries [U(1)
and SO(2)] when $N$ and ${\cal M}$ are conserved as in current
experiments. We discuss the implications to the quantum dynamics
due to an external (homogeneous) magnetic field.
We also comment on the case of a spinor-1 condensate
with anti-ferromagnetic interactions.
\end{abstract}

\maketitle

\section{Introduction}\label{intro}
Since the observation of Bose-Einstein condensation in trapped
dilute alkali atoms \cite{bec,mit,rice}, the coherence properties
of the condensate has become the focus of many theoretical studies
\cite{lewenstein,wright,imamoglu,villain97,law,villain99}. As was
initially pointed out in Refs. \cite{lewenstein,wright}, the
number fluctuations associated with the assumption of a coherent
state macroscopic condensate wavefunction leads to the
``diffusion" (or spreading) of the initial phase. For a scalar,
one component atomic condensate, this diffusion, the attempt to
restore the $U(1)$ symmetry of the underline interacting system,
can be formulated in terms of the dynamics of a zero mode, or the
Goldstone mode \cite{lewenstein}. More physically meaningful
discussions in terms of the relative phases of two condensates
were studied soon afterwards
\cite{imamoglu,villain97,law,villain99}. Starting
with the remarkable direct observation of the first order
coherence in an interference experiment \cite{mitint}, direct
relationship between the number fluctuation and quantum phase
dynamics was first observed with a condensate in a periodic
potential \cite{kasevich}, and more recently, in the remarkable
Mott insulating state \cite{hansch,zoller}.

The emergence of spinor condensates \cite{kurn,mike}
(of atoms with hyperfine quantum number $F=1$) has created
new opportunities to understand quantum coherence and the
associated phase dynamics within a three component condensate
\cite{burnett,ashab}.
In an earlier paper \cite{yil},
we have presented important physical insight into
quantum phase diffusions of a spinor condensate.
We have investigated the coupled zero mode dynamics as a consequence
of the usual mean field treatment that calls for the breaking
of two continuous symmetries: a U(1) gauge transformation
$e^{i\theta}$ and SO(3) spin rotations ${\cal
U}(\alpha,\beta,\tau)=e^{-iF_z\alpha}e^{-iF_y\beta}e^{-iF_z\tau}$
(in the absence of an external magnetic field) \cite{ho}.

This paper provides a detailed investigation
of the quantum phase dynamics for a spinor-1 condensate.
As before, we will focus on the case of ferromagnetic
interactions, for which the validity of
the single spatial mode approximation proves to be a convenient
starting point for the emergence of transparent physical
illustrations of the zero mode dynamics \cite{su}. The fluctuations
of atom numbers in each of the spin states will be shown to be
connected directly, in this case, to the fluctuations of the
pointing direction of the macroscopic spin (or magnetic dipole
moment) of the condensate. This paper is organized as follows. In
section \ref{model}, after introducing the Hamiltonian for a
spinor-1 condensate, we explicitly work out the Heisenberg
equations for the atomic field operators. Following the standard
Bogoliubov approach \cite{bog}, we obtain the coupled
Gross-Pitaevskii equations for the mean field of the condensate
and the coupled Bogoliubov-de Gennes equations for the quantum
fluctuations. We then generalize the Hermitian operators for the
condensate number and phase fluctuations (of the Goldstone mode)
as introduced earlier \cite{lewenstein} to each of the spin
component, and derive their dynamic equations under the rotating
wave function approximation. In section \ref{stead}, we perform
a detailed study of the stationary phase diffusion of a spinor-1
condensate within the single mode approximation valid exactly for
ferromagnetic interactions. Section IV addresses the influence
of an external magnetic field, and in section V we discuss the
same phase dynamics for a condensate with anti-ferromagnetic interactions.
We conclude in section \ref{con}.

\section{Formulation}\label{model}
We consider a system of $N$ spin-1 bosonic atoms interacting via
only s-wave scattering \cite{ho,ohmi,goldstein,pu}. In the
second-quantized form, the Hamiltonian of our system becomes
\cite{ho,ohmi}
\begin{eqnarray}
H&=&\sum_i\int d\,\vec r\psi_i^\dag(\vec
r)\left[-\frac{\hbar^2\nabla^2}{2M}
+V_{\rm ext}(\vec r)\right]\psi_i(\vec r)\nonumber\\
&&+\frac{c_0}{2}\sum_{i,j}\int d\,\vec r\psi_i^\dag(\vec
r)\psi_j^\dag(\vec r)
\psi_i(\vec r)\psi_j(\vec r)\nonumber\\
&&+\frac{c_2}{2}\sum_{i,j,k,l}\int d\,\vec r\psi_i^\dag(\vec r)
\psi_j^\dag(\vec r)\vec{\mathbf F}_{ik} \cdot\vec{\mathbf
F}_{jl}\psi_l(\vec r)\psi_k(\vec r),\hskip 18pt \label{ham}
\end{eqnarray}
where $\psi_j(\vec r)$ ($j=+,0,-$) denotes the annihilation
operator for the $j$-th component of a spinor-1 field. The
external trapping potential $V_{\rm ext}(\vec r)$ is assumed
spin-independent as in a far off-resonant optical dipole force
trap (FORT) so that atomic spin degrees of freedom are
completely accessible. The pair interaction coefficients are
$c_0=4\pi\hbar^2(a_0+2a_2)/3M$ and $c_2=4\pi\hbar^2(a_2-a_0)/3M$,
with $a_0$ ($a_2$) the s-wave scattering length for two spin-1
atoms in the combined symmetric channel of total spin $0$ ($2$),
and $M$ is the mass of atom. $\vec\mathbf F$ is spin 1 matrix
representation. The interacting terms can be regrouped to show
that they include respectively self-scattering, cross-scattering,
and the spin-relaxation \cite{pu}.

From the Hamiltonian (\ref{ham}), one can easily derive the
Heisenberg equations for the field operators
\begin{eqnarray}
i\hbar\dot{\psi}_+&=&\left[-\frac{\hbar^2\nabla^2}{2M}+V_{\rm
ext}(\vec r)\right]\psi_++c_2\psi_-^\dag\psi_0^2\nonumber\\
&&+c_0(\psi_+^\dag\psi_++\psi_0^\dag\psi_0+\psi_-^\dag\psi_-)\psi_+\nonumber\\
&&+c_2(\psi_+^\dag\psi_++\psi_0^\dag\psi_0-\psi_-^\dag\psi_-)\psi_+,\nonumber\\
i\hbar\dot{\psi}_0&=&\left[-\frac{\hbar^2\nabla^2}{2M}+V_{\rm
ext}(\vec r)\right]\psi_0+2c_2\psi_0^\dag\psi_+\psi_-\nonumber\\
&&+c_0(\psi_+^\dag\psi_++\psi_0^\dag\psi_0+\psi_-^\dag\psi_-)\psi_0\nonumber\\
&&+c_2(\psi_+^\dag\psi_++\psi_-^\dag\psi_-)\psi_0,\nonumber\\
i\hbar\dot{\psi}_-&=&\left[-\frac{\hbar^2\nabla^2}{2M}+V_{\rm
ext}(\vec r)\right]\psi_-+c_2\psi_+^\dag\psi_0^2\nonumber\\
&&+c_0(\psi_+^\dag\psi_++\psi_0^\dag\psi_0+\psi_-^\dag\psi_-)\psi_-\nonumber\\
&&+c_2(\psi_-^\dag\psi_-+\psi_0^\dag\psi_0-\psi_+^\dag\psi_+)\psi_-.
\label{hei}
\end{eqnarray}

Following the Hartree mean-field theory we assume there are three
`large' condensate components $\phi_j$ around which we can study
the small quantum fluctuations (off-condensate excitations) via
the Bogoluibov approximation
\begin{eqnarray}
\psi_j(\vec r)=\sqrt{N}\phi_j(\vec r)+\delta\psi_j(\vec r),
\label{bog}
\end{eqnarray}
with $N$ the total number of atoms. We note that such an approach
can become questionable in a multi-component system where certain
component becomes negligibly small and its quantum nature
becomes important. For ferromagnetic interactions, however, our
previous studies \cite{su} show that the ground state wave
functions can always be expressed as $\phi_j\propto\phi$ and hence
the Bogoluibov prescription can always be applied. By substituting
Eq. (\ref{bog}) into (\ref{hei}), we find in the zeroth order, the
coupled Gross-Pitaevskii equations (GPEs),
\begin{eqnarray}
i\hbar\dot{\phi}_+&=&\left[{\cal
L}+c_0\rho+c_2(\rho_++\rho_0-\rho_-)\right]\phi_++c_2N\phi_-^{\ast}\phi_0^2,\nonumber\\
i\hbar\dot{\phi}_0&=&\left[{\cal
L}+c_0\rho+c_2(\rho_++\rho_-)\right]\phi_0+2c_2N\phi_0^{\ast}\phi_+\phi_-,\nonumber\\
i\hbar\dot{\phi}_-&=&\left[{\cal
L}+c_0\rho+c_2(\rho_-+\rho_0-\rho_+)\right]\phi_-+c_2N\phi_+^{\ast}\phi_0^2,
\nonumber\\
\label{gpe3}
\end{eqnarray}
with ${\cal L}=-\hbar^2\nabla^2/2M+V_{\rm ext}(\vec r)$, the
$j$-th component condensate density $\rho_j(\vec r)=N|\phi_j(\vec
r)|^2$, and the total condensate density $\rho(\vec r)=\rho_+(\vec
r)+\rho_0(\vec r)+\rho_-(\vec r)$. $\int d\,\vec r\rho_j(\vec
r)=N_j$ is the number of condensed atoms in the $j$-th component.
In the first order, the quantum fluctuations obey the following
set of Bogoluibov-de Gennes equations (BdGEs),
\begin{eqnarray}
i\hbar\frac{\partial\delta\psi_+}{\partial t}&=&\left[{\cal
L}+c_0(\rho+\rho_+)+c_2(2\rho_++\rho_0-\rho_-)\right]\delta\psi_+\nonumber\\
&&+N(c_+\phi_0^*\phi_++2c_2\phi_-^*\phi_0)\delta\psi_0\nonumber\\
&&+c_-N\phi_-^*\phi_+\delta\psi_-
+c_+N\phi_+^2\delta\psi_+^\dag\nonumber\\
&&+c_+N\phi_+\phi_0\delta\psi_0^\dag
+N(c_-\phi_+\phi_-+c_2\phi_0^2)\delta\psi_-^\dag, \nonumber\\
%%%%%%%%%%%%%%%%%%%%%%%%%%%%%%%%%%%%%%%%%%%%%%%
i\hbar\frac{\partial\delta\psi_0}{\partial t}&=&\left[{\cal
L}+c_0(\rho+\rho_0)+c_2(\rho_++\rho_-)\right]\delta\psi_0\nonumber\\
&&+N(c_+\phi_+^*\phi_0+2c_2\phi_0^*\phi_-)\delta\psi_+\nonumber\\
&&+N(c_+\phi_-^*\phi_0+2c_2\phi_0^*\phi_+)\delta\psi_-\nonumber\\
&&+c_+N\phi_+\phi_0\delta\psi_+^\dag
+c_+N\phi_0\phi_-\delta\psi_-^\dag\nonumber\\
&&+N(c_0\phi_0^2+2c_2\phi_+\phi_-)\delta\psi_0^\dag,\nonumber\\
%%%%%%%%%%%%%%%%%%%%%%%%%%%%%%%%%%%%%%%%%%%%%%%
i\hbar\frac{\partial\delta\psi_-}{\partial t}&=&\left[{\cal
L}+c_0(\rho+\rho_-)+c_2(2\rho_-+\rho_0-\rho_+)\right]\delta\psi_-\nonumber\\
&&+N(c_+\phi_0^*\phi_-+2c_2\phi_+^*\phi_0)\delta\psi_0\nonumber\\
&&+c_-N\phi_+^*\phi_-\delta\psi_++c_+N\phi_-^2\delta\psi_-^\dag\nonumber\\
&&+c_+N\phi_-\phi_0\delta\psi_0^\dag
+N(c_-\phi_+\phi_-+c_2\phi_0^2)\delta\psi_+^\dag,\nonumber\\
\label{fluc}
\end{eqnarray}
where we have defined $c_\pm\equiv c_0\pm c_2$.

To study the quantum phase dynamics, we
introduce the number fluctuation operators \cite{lewenstein}
\begin{eqnarray}
P_j=\int d\,\vec r\left[\phi_j^\ast(\vec r)\delta\psi_j(\vec
r)+\phi_j(\vec r)\delta\psi_j^\dag(\vec r)\right].
\end{eqnarray}
Using the GPEs (\ref{gpe3}) and the BdGEs (\ref{fluc}) above,
we find the evolution equations for these operators,
\begin{eqnarray}
i\hbar\frac{\partial P_+}{\partial t}
&=&-\frac{i\hbar}{2}\frac{\partial P_0}{\partial t}
=i\hbar\frac{\partial P_-}{\partial t}\nonumber\\
&=&c_2N\int d\,\vec r(\phi_-^\ast\phi_0^2\delta\psi_+^\dag
+\phi_+^\ast\phi_0^2\delta\psi_-^\dag\nonumber\\
&&\qquad+2\phi_+^\ast\phi_-^\ast\phi_0\delta\psi_0-h.\,c.).
\end{eqnarray}
Thus both $P_{\rm tot}=P_++P_0+P_-$ and $P_+-P_-$ are constants
of motion, which are in fact obvious since
the Hamiltonian (\ref{ham}) conserves the total
number of atoms and the total magnetization.

We define
the conjugate phase operator $Q_{\rm tot}=Q_++Q_0+Q_-$
according to
\begin{eqnarray}
Q_j=i\hbar\int d\,\vec r\left[\theta_j^\ast(\vec
r)\delta\psi_j(\vec r)-\theta_j(\vec r)\delta\psi_j^\dag(\vec
r)\right].
\end{eqnarray}
The canonical quantization condition $[Q_{\rm tot},P_{\rm
tot}]=i\hbar$ is satisfied if the constraint $J_++J_0+J_-\equiv 1$
is enforced with $J_j\equiv\int d\,\vec r\left[\theta_j^\ast(\vec
r)\phi_j(\vec r) +\theta_j(\vec r)\phi_j^\ast(\vec r)\right]$.
From the defining equation for $Q_{\rm tot}$,
\begin{eqnarray}
\frac{dQ_{\rm tot}}{dt}=N\tilde{u}P_{\rm tot},
\end{eqnarray}
the Goldstone mode inertial
parameter $\tilde{u}$ can be determined.
We obtain the dynamic equations for the phase functions as,
\begin{eqnarray}
i\hbar\dot{\theta}_+&=&\left[{\cal L}+c_0(\rho+\rho_+)
+c_2(2\rho_++\rho_0-\rho_-)\right]\theta_+\nonumber\\
&&+N(c_+\phi_0^*\phi_++2c_2\phi_-^*\phi_0)\theta_0
+c_-N\phi_-^*\phi_+\theta_-\nonumber\\
&&+c_+N\phi_+^2\theta_+^\ast
+c_+N\phi_+\phi_0\theta_0^\ast\nonumber\\
&&+N(c_-\phi_+\phi_-+c_2\phi_0^2)\theta_-^\ast
-N\tilde{u}\phi_+, \label{phase3a}\nonumber\\
%%%%%%%%%%%%%%%%%%%%%%%%%%%%%%%%%%%%%%%%%%%%%%%%%%%%%%%%%%%%%%%%
i\hbar\dot{\theta}_0&=&\left[{\cal L}
+c_0(\rho+\rho_0)+c_2(\rho_++\rho_-)\right]\theta_0\nonumber\\
&&+N(c_+\phi_+^*\phi_0+2c_2\phi_0^*\phi_-)\theta_+\nonumber\\
&&+N(c_+\phi_-^*\phi_0+2c_2\phi_0^*\phi_+)\theta_-\nonumber\\
&&+c_+N\phi_+\phi_0\theta_+^\ast
+c_+N\phi_0\phi_-\theta_-^\ast\nonumber\\
&&+N(c_0\phi_0^2+2c_2\phi_+\phi_-)\theta_0^\ast-N\tilde{u}\phi_0,\label{phase3b}\nonumber\\
%%%%%%%%%%%%%%%%%%%%%%%%%%%%%%%%%%%%%%%%%%%%%%%%%%%%%%%%%%%%%%%%
i\hbar\dot{\theta}_-&=&\left[{\cal L}
+c_0(\rho+\rho_-)+c_2(2\rho_-+\rho_0-\rho_+)\right]\theta_-\nonumber\\
&&+N(c_+\phi_0^*\phi_-+2c_2\phi_+^*\phi_0)\theta_0
+c_-N\phi_+^*\phi_-\theta_+\nonumber\\
&&+c_+N\phi_-^2\theta_-^\ast+c_+N\phi_-\phi_0\theta_0^\ast\nonumber\\
&&+N(c_-\phi_+\phi_-+c_2\phi_0^2)\theta_+^\ast-N\tilde{u}\phi_-.
\label{phase3c}
\end{eqnarray}
Combing Eqs. (\ref{fluc}) and (\ref{phase3c}), the dynamic equations
of $Q_j$ can be easily obtained to be
\begin{widetext}
\begin{eqnarray}
\dot{Q}_+&=&N\tilde{u}P_++N\int d\,\vec r\left\{-\left[
\left(c_+\phi_+^*\phi_0+2c_2\phi_0^*\phi_-\right)\theta_0^*
+c_+\phi_+^*\phi_0^*\theta_0
+\left(c_-\phi_+^*\phi_-^*+c_2\phi_0^{*2}\right)\theta_-
+c_-\phi_+^*\phi_-\theta_-^*\right]\delta\psi_+\right.\nonumber\\
&&+\left.\left(c_+\phi_0^*\phi_+\theta_+^*
+c_+\phi_+^*\phi_0^*\theta_+
+2c_2\phi_-^*\phi_0\theta_+^*\right)\delta\psi_0+\left(
c_-\phi_+^*\phi_-^*\theta_+ +c_-\phi_-^*\phi_+\theta_+^*
+c_2\phi_0^{*2}\theta_+\right)\delta\psi_-+h.c\right\},\nonumber\\
%%%%%%%%%%%%%%%%%%%%%%%%%%%%%%%%%%%%%%%%%%%%%%%%%%%%%%%%%%%%%%%%
\dot{Q}_0&=&N\tilde{u}P_0+N\int d\,\vec r\left\{-\left[
\left(c_+\phi_0^*\phi_++2c_2\phi_-^*\phi_0\right)\theta_+^*
+c_+\phi_+^*\phi_0^*\theta_+
+\left(c_+\phi_0^*\phi_-+2c_2\phi_+^*\phi_0\right)\theta_-^*
+c_+\phi_0^*\phi_-^*\theta_-\right]\delta\psi_0\right.\nonumber\\
&&+\left.\left(c_+\phi_+^*\phi_0\theta_0^*
+c_+\phi_+^*\phi_0^*\theta_0
+2c_2\phi_0^*\phi_-\theta_0^*\right)\delta\psi_+
+\left(c_+\phi_-^*\phi_0\theta_0^* +c_+\phi_0^*\phi_-^*\theta_0
+2c_2\phi_0^*\phi_+\theta_0^*\right)\delta\psi_-+h.c\right\},\nonumber\\
%%%%%%%%%%%%%%%%%%%%%%%%%%%%%%%%%%%%%%%%%%%%%%%%%%%%%%%%%%%%%%%%
\dot{Q}_-&=&N\tilde{u}P_-+N\int d\,\vec r\left\{-\left[
\left(c_+\phi_-^*\phi_0+2c_2\phi_0^*\phi_+\right)\theta_0^*
+c_+\phi_-^*\phi_0^*\theta_0
+\left(c_-\phi_+^*\phi_-^*+c_2\phi_0^{*2}\right)\theta_+
+c_-\phi_-^*\phi_+\theta_+^*\right]\delta\psi_-\right.\nonumber\\
&&+\left.\left(c_+\phi_0^*\phi_-\theta_-^*
+c_+\phi_-^*\phi_0^*\theta_-
+2c_2\phi_+^*\phi_0\theta_-^*\right)\delta\psi_0
+\left(c_-\phi_+^*\phi_-^*\theta_- +c_-\phi_+^*\phi_-\theta_-^*
+c_2\phi_0^{*2}\theta_-\right)\delta\psi_++h.c\right\}.
\end{eqnarray}

We note that $[Q_j,P_k]=i\hbar\delta_{jk}J_j$, $[\delta\psi_j(\vec
r),P_k]=\delta_{jk}\phi_j(\vec r)$, and $[\delta\psi_j(\vec
r),Q_k]=-i\hbar\delta_{jk}\theta_j(\vec r)$. This prompts us to
make the rotating wave approximation (RWA)
\cite{villain97,villain99} by assuming that $\delta\psi_j(\vec
r)=\sum_ja_j(\vec r)P_j+\sum_jb_j(\vec r)Q_j$, which leads to
$a_j(\vec r)=\theta_j(\vec r)/J_j$ and $b_j(\vec r)=\phi_j(\vec
r)/i\hbar J_j$. Therefore, under the RWA, we obtain
$\delta\psi_j(\vec r)=\theta_j(\vec r)P_j/J_j+\phi_j(\vec
r)Q_j/i\hbar J_j$. With this result, the complete dynamic
equations for the number and phase fluctuation operators become
\begin{eqnarray}
\dot{P}_+&=&-\frac{\dot{P}_0}{2}=\dot{P}_-\nonumber\\
&=&\frac{2c_2N}{\hbar}\left[\frac{(\phi_-\theta_+|\phi_0^2)^{\prime\prime}}{J_+}P_+
-2\frac{(\phi_0\theta_0|\phi_+\phi_-)^{\prime\prime}}{J_0}P_0
+\frac{(\phi_+\theta_-|\phi_0^2)^{\prime\prime}}{J_-}P_-\right]\nonumber\\
&&+\frac{2c_2N}{\hbar^2}(\phi_+\phi_-|\phi_0^2)^{\prime}
\left(\frac{Q_+}{J_+}-\frac{2Q_0}{J_0}
+\frac{Q_-}{J_-}\right),\nonumber\\
%%%%%second equation
\dot{Q}_+&=&N\tilde{u}P_+-\frac{2N}{J_+}\left[2c_-I_{-+}+2c_+I_{+0}+
2c_2(\phi_0\theta_0|\phi_-\theta_+)^{\prime}
+c_2(\phi_0^2|\theta_-\theta_+)^{\prime}\right]P_+\nonumber\\
&&+\frac{4N}{J_0}\left[c_+I_{+0}+
c_2(\phi_0\theta_0|\phi_-\theta_+)^{\prime}\right]P_0
+\frac{2N}{J_-}\left[2c_-I_{-+}
+c_2(\phi_0^2|\theta_+\theta_-)^{\prime}\right]P_-
\nonumber\\
&&+\frac{2c_2N}{\hbar J_+} \left[2(\phi_+\phi_-|\phi_0\theta_0)''+
(\phi_+\theta_-|\phi_0^2)''\right]Q_+ +\frac{4c_2N}{\hbar
J_0}(\phi_-\theta_+|\phi_0^2)^{\prime\prime}Q_0
+\frac{2c_2N}{\hbar J_-}(\phi_0^2|\theta_+\phi_-)^{\prime\prime}Q_-,\nonumber\\
%%%%%second equation
\dot{Q}_0&=&N\tilde{u}P_0-\frac{4N}{J_0}\left[c_+I_{+0}+c_+I_{-0}+
c_2(\phi_-\theta_+|\phi_0\theta_0)^{\prime}
+c_2(\phi_+\theta_-|\phi_0\theta_0)^{\prime}\right]P_0\nonumber\\
&&+\frac{4N}{J_+}\left[c_+I_{+0}+
c_2(\phi_0\theta_0|\phi_-\theta_+)^{\prime}\right]P_+
+\frac{4N}{J_-}\left[c_+I_{-0}+
c_2(\phi_0\theta_0|\phi_+\theta_-)^{\prime}\right]P_-\nonumber\\
&&+\frac{4c_2N}{\hbar J_+} (\phi_0\theta_0|\phi_+\phi_-)''Q_+
+\frac{4c_2N}{\hbar J_0}[(\phi_0^2|\phi_-\theta_+)''
+(\phi_0^2|\phi_+\theta_-)'']Q_0 +\frac{4c_2N}{\hbar J_-}
(\phi_0\theta_0|\phi_+\phi_-)''Q_-,\nonumber\\
%%%%%third equation
\dot{Q}_-&=&N\tilde{u}P_--\frac{2N}{J_-}\left[2c_-I_{-+}+2c_+I_{-0}+
2c_2(\phi_0\theta_0|\phi_+\theta_-)^{\prime}
+c_2(\phi_0^2|\theta_+\theta_-)^{\prime}\right]P_-\nonumber\\
&&+\frac{4N}{J_0}\left[c_+I_{-0}+
c_2(\phi_0\theta_0|\phi_+\theta_-)^{\prime}\right]P_0
+\frac{2N}{J_+}\left[2c_-I_{-+}
+c_2(\phi_0^2|\theta_-\theta_+)^{\prime}\right]P_+
\nonumber\\
&&+\frac{2c_2N}{\hbar
J_+}(\phi_0^2|\theta_-\phi_+)^{\prime\prime}Q_+
+\frac{4c_2N}{\hbar
J_0}(\phi_+\theta_-|\phi_0^2)^{\prime\prime}Q_0
+\frac{2c_2N}{\hbar J_-} \left[2(\phi_-\phi_+|\phi_0\theta_0)''+
(\phi_-\theta_+|\phi_0^2)''\right]Q_-, \label{dynpq}
\end{eqnarray}
\end{widetext}
where we have introduced the following shorthand notation;
$(fg|hy)=\int d\,\vec rf^\ast(\vec r)g^\ast(\vec r)h(\vec r)y(\vec
r)$, $(.)^{\prime}={\rm Re}(.)$, $(.)^{\prime\prime}={\rm Im}(.)$,
and $I_{ml}=\int d\,\vec r\left(\phi_m\theta_m^\ast\right)'
\left(\phi_l\theta_l^\ast\right)'$. Equation (\ref{dynpq}) thus
completely determines the zero mode condensate fluctuations.
Before finding the corresponding compact zero mode Hamiltonian in
quadratic forms of $P_j$ and $Q_j$, we will first attempt to
simplify in the next section using general properties of
$\phi_j(\vec r)$ and $\theta_j(\vec r)$ for a ferromagnetic
spinor-1 condensate \cite{su}.

\section{Phase diffusions and the zero modes}\label{stead}
In general, the ground state
wave functions of a spinor condensate can be expressed as
\begin{eqnarray}
\phi_j(\vec r,t)=|\phi_j(\vec r)| e^{-i\mu
t/\hbar+i\alpha_j},\label{pha}
\end{eqnarray}
with a common chemical potential $\mu$ for all spin components
when the external magnetic field is negligible. Based on
the minimization for the total energy Eq. (\ref{ham})
within the mean field theory, it was
shown in Refs. \cite{isoshima,robins} that there exists an important
relation among $\alpha_j$, given by  $\alpha_++\alpha_--2\alpha_0 =0$.
We further proved in Ref. \cite{su} that the ground state wave
function for each of the spin component takes the same
spatial shape for (ferromagnetic interactions), i.e.
\begin{eqnarray}
\phi_j(\vec r)=\sqrt{n_j}\,\phi(\vec r)e^{i\alpha_j}, \label{wave}
\end{eqnarray}
with the real-valued mode function $\phi(\vec r)$ (normalized to
unity) governed by an equivalent scalar condensate GPE
\begin{eqnarray}
\left[{\cal L}+c_+N\phi^2(\vec r)\right]\phi(\vec r) =\mu\phi(\vec
r), \label{gpe1}
\end{eqnarray}
of a scattering length $a_2$ (note $c_+\propto a_2$). Atoms thus
only collide in the symmetric total spin $F=2$ channel in a
ferromagnetic state in order to maintain their individual spins
parallel. $n_j=N_j/N$ is the ratio of the number of atoms in the
$i$-th spin component to the total number of atoms. For any given
magnetization ${\cal M}$, $n_j$ is given explicitly by
$n_\pm=(1\pm m)^2/4$ and $n_0=(1-m^2)/2$ with $m={\cal M}/N$
\cite{pu,su}.

Similarly, we look for phase functions of the form
\begin{eqnarray}
\theta_j(\vec r,t)=|\theta_j(\vec r)|e^{-i\mu_j t/\hbar+i\alpha_j}.
\label{th}
\end{eqnarray}
Substituting Eq. (\ref{th}) into Eq. (\ref{phase3c}),
we find that $\theta_j(\vec r,t)$ also evolve
in time as $e^{-i\mu t/\hbar}$. To focus on the steady state quantum
fluctuation properties, we may therefore neglect the
time-dependent part in $\phi_j(\vec r,t)$ and $\theta_j(\vec r,t)$
from now on. The steady state equations for phase functions now
become
\begin{eqnarray}
\left[{\cal L}-\mu+c_0(\rho+2\rho_+)
+c_2(3\rho_++\rho_0-\rho_-)\right]\xi_+\nonumber\\
+2N(c_+\phi_++c_2\phi_-)\phi_0\xi_0
+N(2c_-\phi_+\phi_-+c_2\phi_0^2)\xi_-\nonumber\\
=N\tilde{u}\phi_+,\nonumber\\
%%%%%%%%%%%%%%%%%%%%%%%%%%%%%%%%%%%%%%%%%%%%%%%%%%
\left[{\cal L}-\mu+c_0(\rho+2\rho_0)
+c_2(\rho_++2N\phi_+\phi_-+\rho_-)\right]\xi_0\nonumber\\
+2N(c_+\phi_++c_2\phi_-)\phi_0\xi_+
+2N(c_+\phi_-+c_2\phi_+)\phi_0\xi_-\nonumber\\
=N\tilde{u}\phi_0,\nonumber\\
%%%%%%%%%%%%%%%%%%%%%%%%%%%%%%%%%%%%%%%%%%%%%%%%%%
\left[{\cal L}-\mu+c_0(\rho+2\rho_-)
+c_2(3\rho_-+\rho_0-\rho_+)\right]\xi_-\nonumber\\
+2N(c_+\phi_-+c_2\phi_+)\phi_0\xi_0
+N(2c_-\phi_+\phi_-+c_2\phi_0^2)\xi_+\nonumber\\
=N\tilde{u}\phi_-,\hskip 12pt
\label{eqstp}
\end{eqnarray}
where $\xi_j(\vec r)=\theta_j(\vec r)e^{-i\alpha_j}$ are now
real-valued functions. One can easily check that similar to the
single mode approximation for $\phi_j(\vec r)$, $\xi_j(\vec r)$
can also be expressed as $\xi_j(\vec r)=n_j^{1/2}\,\theta(\vec r)$
with $\theta(\vec r)$ governed by the following equation
\begin{eqnarray}
\left[{\cal L}-\mu+3c_+N\phi^2(\vec r)\right]\theta(\vec r)
=N\tilde{u}\phi(\vec r), \label{eqphase}
\end{eqnarray}
and $\tilde{u}$ is determined through the normalization constraint
$\int d\,\vec r\phi(\vec r)\theta(\vec r)=1/2$. Therefore, for
ferromagnetic interactions, the phase functions can be generally
expressed as
\begin{eqnarray}
\theta_j(\vec r)=\sqrt{n_j}\,\theta(\vec r)e^{i\alpha_j}.
\label{phase}
\end{eqnarray}

We note that Eq. (\ref{eqphase}) is in fact identical to the
equation satisfied by the phase function of a scalar condensate
\cite{lewenstein}, a point easily understood for ferromagnetic
interactions when all spins align along the same direction. If we
take this direction as the quantization axis, then the spinor
condensate is essentially a scalar condensate since its total
magnetization ${\cal M}$ is conserved. All overlap integrals in
Eq. (\ref{dynpq}) are presented in Appendix \ref{over}. They are
all real quantities with the choice of phase parameters.

\begin{figure}[h]
\centering
\includegraphics[width=3in]{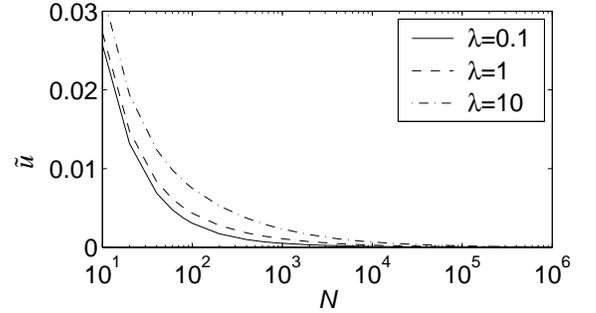}
\caption{The $N$-dependence of $\tilde{u}$ (in units of
$\hbar\omega_\perp$) for a $^{87}$Rb condensate in a cylindrically
symmetric trap with $\omega_x=\omega_y=\omega_\perp=(2\pi) 100$
(Hz) and $\omega_z=\lambda\omega_\perp$.
We have used $a_0=101.8a_B$ and $a_2=100.4a_B$ ($a_B$
Bohr radius) \cite{green}. } \label{figue}
\end{figure}

We now attempt to express the zero mode dynamics in terms of its
associated Hamiltonian. Since $P_j$ and $Q_j$ are not canonically
conjugated variables, we define $p_j'\equiv P_j$ and $q_j'\equiv
Q_j/J_j$ such that they become canonically conjugates.
The zero mode Hamiltonian that governs the dynamics of Eq.
(\ref{dynpq}) is
\begin{eqnarray}
\frac{H_{\rm zero}}{N} &=&\left(\frac{\tilde{u}}{2n_+{\cal
O_{\phi\theta}}} -2c_+\frac{1-n_+}{n_+}-2c_2\frac{n_-}{n_+}\right)
p_+'^2{\cal O}_{\phi\theta}\nonumber\\
&&+\left(\frac{\tilde{u}}{2n_0{\cal O_{\phi\theta}}}
-2c_+\frac{1-n_0}{n_0}-2c_2\right)
p_0'^2{\cal O}_{\phi\theta}\nonumber\\
&&+\left(\frac{\tilde{u}}{2n_-{\cal
O_{\phi\theta}}}-2c_+\frac{1-n_-}{n_-}-2c_2\frac{n_+}{n_-}
\right)p_-'^2{\cal O}_{\phi\theta}\nonumber\\
&&+4\left[c_++c_2\left(\frac{n_-}{n_+}\right)^{1/2}\right]p_+'p_0'
{\cal O}_{\phi\theta}\nonumber\\
&&+4\left[c_++c_2\left(\frac{n_+}{n_-}\right)^{1/2}\right]
p_-'p_0'{\cal O}_{\phi\theta}\nonumber\\
&&+4c_0p_+'p_-'{\cal O}_{\phi\theta}
-\frac{c_2n_0^2{\cal O}_{\phi\phi}}{2\hbar^2}
\left(q_+'-2q_0'+q_-'\right)^2\nonumber\\
&=&{\mathbf p}'^T{\mathbf{\cal A}{\mathbf p}'} +{\mathbf
q}'^T{\mathbf{\cal B}{\mathbf q}'},\label{hzero2}
\end{eqnarray}
where ${\mathbf p}'^T=(p_+',p_0',p_-')$, ${\mathbf
q}'^T=(q_+',q_0',q_-')$, ${\mathbf{\cal A}}$ and ${\mathbf{\cal
B}}$ are two Hermitian matrices whose elements are easily
identified from above equation.
We also present, in Appendix \ref{zero},
the zero mode Hamiltonian in the form of overlap
integrals.
To simplify notations, we have
also defined ${\cal O}_{\phi\theta}\equiv \int d\,\vec
r\phi^2(\vec r)\theta^2(\vec r)$ and ${\cal
O}_{\phi\phi}\equiv\int d\,\vec r\phi^4(\vec r)$.
We find that
matrix $\cal B$ is positive definite since $c_2<0$ for
ferromagnetic spinor condensates. To show the positive
definiteness of $\mathbf{\cal A}$, we decompose $\tilde{u}$ as
\begin{eqnarray}
\tilde{u}=\varepsilon+4c_+{\cal O}_{\phi\theta},\label{utld}
\end{eqnarray}
where
\begin{eqnarray}
\varepsilon=\frac{2}{N}\int d\vec r\theta(\vec r)\left[{\cal L}
-\mu+c_+N\phi^2(\vec r)\right]\theta(\vec r).
\end{eqnarray}
We note that $\varepsilon\equiv 0$ for
noninteracting atoms [$\phi(\vec r)=\theta(\vec r)$].
However, it's nonzero in general and we can prove that
$\varepsilon>0$ since $\theta(\vec r)$ is the solution of Eq.
(\ref{eqphase}). By using Eq. (\ref{utld}), we then obtain
\begin{eqnarray}
{\mathbf p}'^T{\mathbf{\cal A}}{\mathbf p}'&=&
\frac{\varepsilon}{2} \left(\frac{p_+'^2}{n_+}
+\frac{p_0'^2}{n_0}+\frac{p_-'^2}{n_-}\right) +2{\cal
O_{\phi\theta}}{\mathbf p}'^T{\mathbf{\cal A}}'{\mathbf
p}',\nonumber
\end{eqnarray}
where
\begin{eqnarray}
{\mathbf{\cal A}}'=\left(\begin{array}{ccc}
c_0+\frac{4c_2m}{(1+m)^2}&c_0+\frac{2c_2}{1+m}&c_0\\
c_0+\frac{2c_2}{1+m}&c_0&c_0+\frac{2c_2}{1-m}\\
c_0&c_0+\frac{2c_2}{1-m}&c_0-\frac{4c_2m}{(1-m)^2}
\end{array}\right).\nonumber
\end{eqnarray}
It is then simply to verify that $\mathbf{\cal A}'$ is
semi-positive definite, which guarantees that $\mathbf{\cal A}$ is
positive definite. Thus the initial ground state, the mean field
symmetry breaking state with a coherent condensate amplitude
$\sqrt{N}\phi_j(\vec r)$ is stable. The associated quantum
fluctuations of the atom numbers can be studied by the
linearization approximation Eq. (\ref{bog}).

\begin{figure}
\centering
\includegraphics[width=3in]{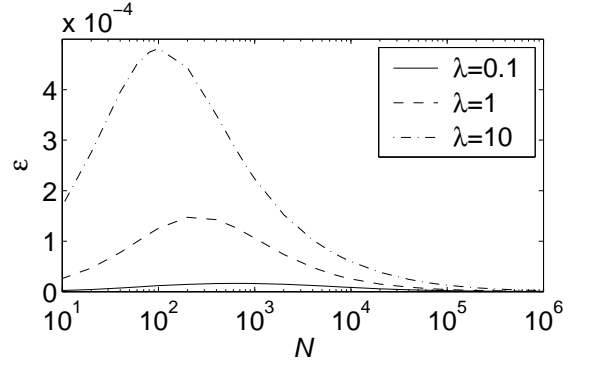}
\caption{The $N$-dependence of $\varepsilon$ (in units of
$\hbar\omega_\perp$) for the same parameters as in Fig.
\ref{figue}.} \label{fige}
\end{figure}

We note that $\mathbf{\cal B}$ is diagonalized
by an orthogonal matrix
\begin{eqnarray}
{\mathbf{\cal U}}=\left(\begin{array}{ccc}
1/\sqrt{3}&1/\sqrt{3}&1/\sqrt{3}\\
1/\sqrt{2}&0&-1/\sqrt{2}\\
1/\sqrt{6}&-2/\sqrt{6}&1/\sqrt{6}
\end{array}\right),
\label{bt}
\end{eqnarray}
which motivates the introduction of new number and phase
fluctuation operators according to ${\mathbf p}^T={\mathbf
p}'^T{\mathbf{\cal U}}^T=(p_N,p_M,p_Y)$ and ${\mathbf
q}^T={\mathbf q}'^T{\mathbf{\cal U}}^T=(q_N,q_M,q_Y)$. Not
surprisingly, this transformation recovers the underline symmetry
of a spinor-1 condensate as discovered recently in
Ref. \cite{ozgur}. It turns out that $p_N$, $p_M$, and $p_Y$
represent respectively the fluctuations of the total number of
atoms, the magnetization, and the hypercharge. $\mathbf{\cal A}$
is not simultaneously diagonalized by the $\mathbf{\cal U}$, yet
the zero mode Hamiltonian takes a much simpler form
\begin{eqnarray}
\frac{H_{\rm zero}}{N}&=&ap_N^2+bp_M^2+cp_Y^2\nonumber\\
&&+\alpha p_Np_Y+\beta p_Mp_Y
+\gamma p_Np_M\nonumber\\
&&+\eta q_Y^2,\label{hzero}
\end{eqnarray}
where all the coefficients are given in Appendix \ref{coeff}.
By replacing $p_j$ with $-i\hbar\partial/\partial q_j$,
we obtain an eigenvalue equation for the distribution
of the various fluctuations in the condensate ground state
\begin{eqnarray}
\frac{E}{N\hbar^2}\varphi&=&-a\frac{\partial^2\varphi}{\partial q_N^2}
-b\frac{\partial^2\varphi}{\partial q_M^2}
-c\frac{\partial^2\varphi}{\partial q_Y^2}\nonumber\\
&&-\alpha\frac{\partial^2\varphi}{\partial q_Nq_Y}
-\beta\frac{\partial^2\varphi}{\partial q_Mq_Y}
-\gamma\frac{\partial^2\varphi}{\partial q_Nq_M}\nonumber\\
&&+\frac{\eta}{\hbar^2} q_Y^2\varphi.
\label{eigenzero}
\end{eqnarray}
We look for eigenfunctions of the form
\begin{eqnarray}
\chi(q_Y)e^{i(k_Nq_N+k_Mq_M)},
\label{sol}
\end{eqnarray}
which upon substituting into (\ref{eigenzero}) yields
\begin{eqnarray}
c\frac{d^2\chi}{dq_Y^2}+i\kappa\frac{d\chi}{dq_Y}
+\left(E'-\frac{\eta}{\hbar^2}q_Y^2\right)\chi=0,
\label{eigeneqn}
\end{eqnarray}
with
\begin{eqnarray}
\kappa &=& \alpha k_N+\beta k_M,\nonumber\\
E' &=& \frac{E}{N\hbar^2}-(ak_N^2+bk_M^2+\gamma k_Nk_M).
\end{eqnarray}
The eigen-equation (\ref{eigeneqn}) is essentially
of the harmonic type, which can be easily solved and
the resulting eigenvalues of Eq. (\ref{eigenzero}) are
\begin{eqnarray}
\frac{E_n}{N}=(2n+1)\hbar(c\eta)^{1/2}+\Lambda(k_N,k_M),\nonumber
\end{eqnarray}
where $n=0,1,2,\ldots$ and
\begin{eqnarray}
\Lambda(k_N,k_M)&=&\hbar^2(ak_N^2+bk_M^2+\gamma k_Nk_M)\nonumber\\
&&-\frac{\hbar^2}{4c}(\alpha k_N+\beta k_M)^2.\nonumber
\end{eqnarray}
We note that $\Lambda(k_N,k_M)$ must be greater than or equal to
zero due to the positive definiteness of matrix ${\mathbf{\cal A}
}$. Therefore for the ground state of Eq. (\ref{eigenzero}), we
must have $k_N=k_M=0$, the ground state ($n=0$) energy and
eigenfunction are respectively
\begin{eqnarray}
E_0&=&N\hbar(c\eta)^{1/2}\equiv \frac{1}{2}\hbar\Omega,\\
\varphi_0(q_Y)&=&\pi^{-1/4}(\hbar\Delta_{Y0})^{-1/2}
\exp\left(-\frac{q_Y^2}{2\hbar^2\Delta_{Y0}^2}\right),
\end{eqnarray}
with
$$\Omega=\frac{2N}{\hbar}[-c_2{\cal O}_{\phi\phi}(\varepsilon-4c_2{\cal O}_{\phi\theta})]^{1/2},$$ and
$$\Delta_{Y0}^2=\frac{1}{\hbar}\sqrt{\frac{c}{\eta}}
=\frac{1}{3n_0^2} \left(\frac{\varepsilon-4c_2{\cal
O}_{\phi\theta}} {-c_2{\cal O}_{\phi\phi}}\right)^{1/2}.$$

\begin{figure}
\centering
\includegraphics[width=3in]{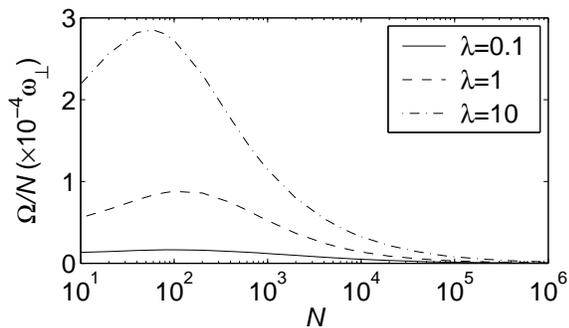}
\caption{The $N$-dependence of $\Omega/N$ for the same parameters
as in Fig. \ref{figue}.} \label{figod}
\end{figure}

\begin{figure}
\centering
\includegraphics[width=3.2in]{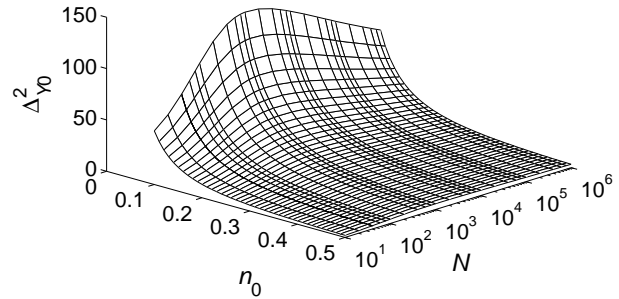}
\caption{The width $\Delta_{Y0}^2$ of the ground state
distribution for $q_Y$ as a function of $n_0$ and $N$ for the
same parameters as in Fig. \ref{figue} and $\lambda=1$.}
\label{figdelta}
\end{figure}

This ground state of $H_{\rm zero}$ can be easily understood. For
a system starting with a fixed total number of atoms and
magnetization, the dynamic conservation of the total number of
atoms and total magnetization as dictated by the Hamiltonian
(\ref{ham}) requires $\langle p_N^2\rangle= \langle
p_M^2\rangle=0$, which leads to $\langle q_N^2\rangle= \langle
q_M^2\rangle=\infty$, i.e., the ground state of $H_{\rm zero}$ has
completely diffused phases for $q_N$ and $q_M$; The inherent
ferromagnetic interaction, nevertheless, prepares a correlated
ground state such that both $p_Y$ and $q_Y$ takes a Gaussian form
distribution with $\langle p_Y^2\rangle= 1/2\Delta_{Y0}^2$ and
$\langle q_Y^2\rangle=\hbar^2\Delta_{Y0}^2/2$, which is in fact
the minimal uncertainty coherent state that is consistent with the
symmetry breaking mean field assumptions of the ground state with
three separate macroscopic wave functions for each of the spin
components. However, it is experimentally difficult to produce a
condensate having total fixed number and phase fluctuations as
specified by such a state. We therefore turn to study the
number and phase fluctuations of a general state. We start with
the dynamic equations for $p_j$ and $q_j$,
\begin{eqnarray}
\dot{p}_N&=&0,\nonumber\\
\dot{p}_M&=&0,\nonumber\\
\dot{p}_Y&=&-2N\eta q_Y,\nonumber\\
\dot{q}_N&=&N\left(2ap_N+\gamma p_M+\alpha p_Y\right),\nonumber\\
\dot{q}_M&=&N\left(\gamma p_N+2bp_M+\beta p_Y\right),\nonumber\\
\dot{q}_Y&=&N\left(\alpha p_N+\beta p_M+2cp_Y\right).
\end{eqnarray}
By noting that both $p_N\equiv p_N(0)$ and $p_M\equiv p_M(0)$ are
constants of motion, we can express the solutions as
\begin{eqnarray}
{\mathbf x}(t)={\mathbf{\cal T}}(t){\mathbf x}(0),
\label{trans}
\end{eqnarray}
where ${\mathbf
x}^T(t)=[p_N(t),p_M(t),p_Y(t),q_N(t),q_M(t),q_Y(t)]$,
${\mathbf x}^T(0)=[p_N(0),p_M(0),p_Y(0),q_N(0),q_M(0),q_Y(0)]$ is its
initial value, and the evolution matrix ${\mathbf{\cal T}}(t)$
takes following form
\begin{widetext}
\begin{eqnarray}
{\mathbf {\cal T}}(t)=\left(\begin{array}{cccccc}1&0&0&0&0&0\\0&1&0&0&0&0
\\-\alpha'(1-\cos\Omega t)&-\beta'(1-\cos\Omega t)&\cos\Omega t
&0&0&-\sin\Omega t/\hbar\Delta_{Y0}^2
\\\zeta_aNt+\hbar\Delta_{Y0}^2\alpha'^2\sin\Omega t&
\zeta_cNt+\hbar\Delta_{Y0}^2\alpha'\beta'\sin\Omega t&
\hbar\Delta_{Y0}^2\alpha'\sin\Omega t&1& 0&-\alpha'(1-\cos\Omega
t)
\\\zeta_cNt+\hbar\Delta_{Y0}^2\alpha'\beta'\sin\Omega t&
\zeta_bNt+\hbar\Delta_{Y0}^2\beta'^2\sin\Omega t&
\hbar\Delta_{Y0}^2\beta'\sin\Omega t&0& 1&-\beta'(1-\cos\Omega t)
\\\hbar\Delta_{Y0}^2\alpha'\sin\Omega t&\hbar\Delta_{Y0}^2\beta'\sin\Omega t
&\hbar\Delta_{Y0}^2\sin\Omega t&0&0&\cos\Omega
t\end{array}\right).\label{tr}
\end{eqnarray}
\end{widetext}
The coefficients $\alpha'$, $\beta'$, $\zeta_a$, $\zeta_b$, and
$\zeta_c$ are given in Appendix \ref{tran}. The explicit solutions
of the time-dependent number and phase fluctuations are given in
Appendix \ref{time}. We note that the general solutions has a
simple structure: in addition to oscillating terms of the form
$\cos\Omega t$ and $\sin\Omega t$, the phase fluctuations of $q_N$
and $q_M$ also contains terms proportional to $Nt$. This indicates
that our linearization approximation Eq. (\ref{bog}) is valid only
for a finite duration. The time-dependent covariance can also be
obtained to be
\begin{eqnarray}
\langle x_i(t)x_j(t)\rangle=\sum_{kl}{\cal T}_{ik}(t){\cal
T}_{jl}(t) \langle x_k(0)x_l(0)\rangle.
\end{eqnarray}

Finally, let's consider the diffusion of the direction of the
macroscopic spin $\vec f(\vec r)\equiv\sum_{ij}\psi_j^\dag(\vec r)
\vec{\mathbf F}_{ij}\psi_j(\vec r)$. As we have shown in Ref.
\cite{su}, independent of the spatial coordinates, the spins of
Bose condensed atoms are parallel for ferromagnetic interactions,
i.e., it essentially acts as a macroscopic magnetic dipole pointing
along the same direction,
\begin{eqnarray}
\vec f_0(\vec r)&=&N\sum_{ij}\phi_i^*(\vec r)
\vec{\mathbf F}_{ij}\phi_j(\vec r)\nonumber\\
&=&N\phi^2(\vec r)\left(\begin{array}{c}\sqrt{1-m^2}\cos\Theta\\
-\sqrt{1-m^2}\sin\Theta\\m\end{array}\right),
\end{eqnarray}
where $\Theta=\alpha_+-\alpha_0=\alpha_0-\alpha_-$.
 Our initial choice of three separate
macroscopic condensate wave functions (with given
phases) for each of
the spin components fixes the initial direction
of the condensate spin direction. As the phase
spreading dynamics attempts to restore the $U(1)$
symmetries of each of the phases, our initial choice
of phases will become irrelevant.
Under the linear
approximation (\ref{bog}) the fluctuation becomes
\begin{eqnarray}
&&\delta\vec f(t)=\int d\vec r\left[\vec f(\vec r)-\vec f_0(\vec
r)\right],
\end{eqnarray}
or explicitly
\begin{eqnarray}
\delta f_x&=&\frac{1}{\sqrt{2}}\int d\vec r(
\phi_0^*\delta\psi_++\phi_+\delta\psi_0^\dag
+\phi_-^*\delta\psi_0+\phi_0\delta\psi_-^\dag\nonumber\\
&&\qquad+h.c.),\nonumber\\
%%%%%%%%%%%%%%%%%%%%%%%%%%%%%%%%%%%%%%%%%%%%%%%%%%%%%%%
\delta f_y&=&\frac{i}{\sqrt{2}}\int d\vec r(
\phi_0^*\delta\psi_++\phi_+\delta\psi_0^\dag
+\phi_-^*\delta\psi_0+\phi_0\delta\psi_-^\dag\nonumber\\
&&\qquad-h.c.),\nonumber\\
%%%%%%%%%%%%%%%%%%%%%%%%%%%%%%%%%%%%%%%%%%%%%%%%%%%%%%%
\delta f_z&=&\int d\vec r(\phi_+^*\delta\psi_+
-\phi_-^*\delta\psi_-+h.c.).
\end{eqnarray}
Although $\langle\delta\vec f(t)\rangle\equiv 0$, the dynamics of
the zero mode causes the variance of $\delta\vec f(t)$ to be
nonzero. It can be reexpressed in terms of the (time-dependent)
number and phase fluctuation operators under the RWA as
\begin{eqnarray}
\delta f_x(t)&=&\frac{\cos\Theta}{\sqrt{1-m^2}}\left[
\sqrt{3}p_N(t)-\sqrt{2}mp_M(t)\right]\nonumber\\
&&+\frac{\sqrt{1-m^2}\sin\Theta}{2\hbar}\left[
\sqrt{2}q_M(t)+\sqrt{6}mq_Y(t)\right],\nonumber\\
%%%%%%%%%%%%%%%%%%%%%%%%%%%%%%%%%%%%%%%%%%%%%%%%%%%%%%%
\delta f_y(t)&=&\frac{-\sin\Theta}{\sqrt{1-m^2}}\left[
\sqrt{3}p_N(t)-\sqrt{2}mp_M(t)\right]\nonumber\\
&&+\frac{\sqrt{1-m^2}\cos\Theta}{2\hbar}\left[
\sqrt{2}q_M(t)+\sqrt{6}mq_Y(t)\right],\nonumber\\
%%%%%%%%%%%%%%%%%%%%%%%%%%%%%%%%%%%%%%%%%%%%%%%%%%%%%%%
\delta f_z(t)&=&\sqrt{2}p_M(t),
\label{t34}
\end{eqnarray}
where the time-dependent terms are
\begin{eqnarray}
&&\sqrt{2}q_M(t)+\sqrt{6}mq_Y(t)=\sqrt{2}q_M(0)+\sqrt{6}mq_Y(0)\nonumber\\
&&\qquad+\frac{4\varepsilon
Nt}{1-m^2}\left[-\sqrt{3}mp_N(0)+\sqrt{2}p_M(0)\right].
\label{timedep}
\end{eqnarray}
We see that $\delta f_z$ is a constant of motion due to the
conservation of total magnetization. On the other hand, both
$\delta f_x$ and $\delta f_y$ diffuse with time. To calculate the
variance of $\delta\vec f(t)$, we simply take $\Theta=0$, which
yields
\begin{eqnarray}
\delta f_x(t)&=&\frac{1}{\sqrt{1-m^2}}\left[
\sqrt{3}p_N(0)-\sqrt{2}mp_M(0)\right],\nonumber\\
%%%%%%%%%%%%%%%%%%%%%%%%%%%%%%%%%%%%%%%%%%%%%%%%%%%%%%%
\delta f_y(t)&=&\frac{\sqrt{1-m^2}}{2\hbar}\left[
\sqrt{2}q_M(t)+\sqrt{6}mq_Y(t)\right],\nonumber\\
%%%%%%%%%%%%%%%%%%%%%%%%%%%%%%%%%%%%%%%%%%%%%%%%%%%%%%%
\delta f_z(t)&=&\sqrt{2}p_M(0),
\end{eqnarray}
i.e., with the direction of the initial macroscopic spin
constrained in the x-z plane. The results of the variances
for condensate spin direction fluctuations are listed
in Appendix \ref{car}.

\begin{figure}
\centering
\includegraphics[width=2.5in]{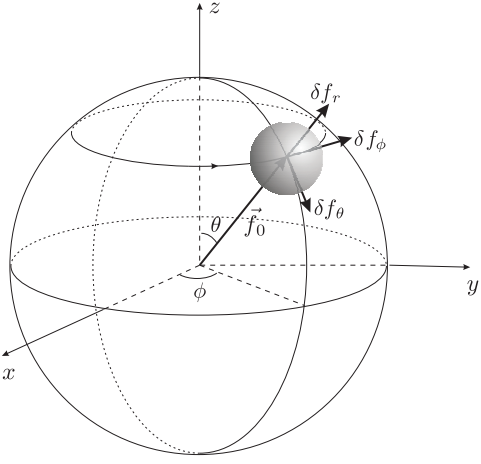}
\caption{Illustration of the macroscopic condensate spin and its
fluctuation in spherical coordinates.} \label{figspin}
\end{figure}

We can also project the fluctuation of the spin [Eq. (\ref{t34})]
onto the spherical coordinates $(\hat r,\hat\theta,\hat\phi)$ as
illustrated in Fig. \ref{figspin}, and noting that
\begin{eqnarray}
\hat r&=&\hat x\sin\theta\cos\phi+\hat y\sin\theta\sin\phi
+\hat z\cos\theta,\nonumber\\
\hat \theta&=&\hat x\cos\theta\cos\phi+\hat y\cos\theta\sin\phi
-\hat z\sin\theta,\nonumber\\
\hat\phi&=&-\hat x\sin\phi+\hat y\cos\phi,
\end{eqnarray}
and
\begin{eqnarray}
\sin\theta=\sqrt{1-m^2},&&\quad \cos\theta=m,\nonumber\\
\sin\phi=-\sin\Theta,\ \ \ &&\quad\cos\phi=\cos\Theta,\nonumber
\end{eqnarray}
we find
\begin{eqnarray}
\delta f_r(t)&=&\sqrt{3}p_N(0),\nonumber\\
\delta f_\theta(t)&=&\frac{1}{\sqrt{1-m^2}}
\left[\sqrt{3}mp_N(0)-\sqrt{2}p_M(0)\right],\nonumber\\
\delta f_\phi(t) &=&\frac{\sqrt{1-m^2}}{2\hbar}
\left[\sqrt{2}q_M(t)+\sqrt{6}mq_Y(t)\right].
\end{eqnarray}
We immediately see that both $\delta f_r(t)\equiv\delta
f_r(0)$ and $\delta f_\theta(t)\equiv\delta f_\theta(0)$ are fixed
due to the conservations of $N$ and ${\cal M}$.
Furthermore, with the use of Eq. (\ref{timedep}),
we find that $\delta f_\phi(t)$ can be rewritten as
\begin{eqnarray}
\delta f_\phi(t)=\delta f_\phi(0)-\frac{2\varepsilon
Nt}{\hbar}\delta f_\theta(0),
\end{eqnarray}
where
$$\delta f_\phi(0)=\frac{\sqrt{1-m^2}}{2\hbar}\left[\sqrt{2}q_M(0)
+\sqrt{6}mq_Y(0)\right].$$
We then see that $\delta
f_\phi(t)$ grows linearly with $t$ and the
diffusion rate is given by
\begin{eqnarray}
R_d=-\frac{2\varepsilon N}{\hbar}\delta f_\theta(0),
\end{eqnarray}
which is proportional to $\delta f_\theta(0)$. A result easily
understandable in terms of the single axis twisting of the isospin
$T_z^2$ (of the Hamiltonian) of a spior-1 spinor condensate
\cite{ozgur}.

Given any initial condensate state, its variances in atom numbers
for each individual spin component and their correlations and
related phase fluctuations completely determine the subsequent
fluctuations of $\delta f_\phi(t)$. $P_j(0)$ is in fact nothing
but the atom number fluctuation of state $|j\rangle$. In general
we we note that
\begin{eqnarray}
\int d\vec r\psi_j^\dag(\vec r)\psi_j(\vec r)&=&N+\sqrt{N}P_j+\int
d\vec r\delta\psi_j^\dag(\vec r)\delta\psi_j(\vec r)\nonumber\\
&=&N+\sqrt{N}P_j\nonumber\\
&&+\frac{s
n_j}{2J_j^2}P_j^2+\frac{n_j}{2\hbar^2J_j^2}Q_j^2-\frac{1}{2},
\end{eqnarray}
where $s=\int d\vec r\,\theta^2(\vec r)$ and we have used RWA in
deriving the last line. One can easily find
\begin{eqnarray}
\delta N_j^2 &=& \langle N_j^2\rangle-\langle N_j\rangle^2\nonumber\\
&=&\left\langle \left[\int d\vec r\psi_j^\dag(\vec r)\psi_j(\vec
r)\right]^2\right\rangle -\left\langle \int d\vec
r\psi_j^\dag(\vec r)\psi_j(\vec r)\right\rangle^2\nonumber\\
&\approx &N\langle P_j^2\rangle=N\langle p'^2_j\rangle,
\end{eqnarray}
to lowest order.

If we use $\sigma_N^2$, $\sigma_M^2$, and $\sigma_Y^2$ to denote
the variances of the total atom number ($N$), the magnetization
(${\cal M}$), and the hyper-charge ($Y$), we find that
\begin{eqnarray}
\left\langle\left[\delta f_r(t)\right]^2\right\rangle&=&3\sigma_N^2,\nonumber\\
\left\langle\left[\delta
f_\theta(t)\right]^2\right\rangle&=&\frac{1}{1-m^2}\left(
3m^2\sigma_N^2+2\sigma_M^2\right).
\end{eqnarray}
To calculate the variance of $\delta f_\phi(t)$,
we assume a completely uncorrelated
(between fluctuations of $N$, ${\cal M}$, and $Y$)
initial distribution in the form of a Gaussian wave packet
\begin{eqnarray}
\varphi(p_N,p_M,p_Y)=(2\pi)^{-3/4}(\sigma_N\sigma_M\sigma_Y)^{-1/2}\nonumber\\
\times\exp\left[-\frac{1}{4}\left(
\frac{p_N^2}{\sigma_N^2}+\frac{p_M^2}{\sigma_M^2}+\frac{p_Y^2}{\sigma_Y^2}
\right)\right].
\label{npj}
\end{eqnarray}
Since $q_j=i\hbar\partial/\partial p_j$, we find
for such a distribution
\begin{eqnarray}
\langle
q_i(0)q_j(0)\rangle &=&{\hbar^2\over 4\sigma_i^2}\delta_{ij},\nonumber\\
\langle p_i(0)p_j(0)\rangle &=&{\sigma_i^2}\delta_{ij},\nonumber\\
\langle q_i(0)p_i(0)+p_i(0)q_i(0)\rangle &=&0,\nonumber\\
\langle q_i(0)p_j(0)\rangle &=&0, \mbox{ for $i\neq j$},
\end{eqnarray}
and $i,j=N,M,Y$. Therefore
\begin{eqnarray}
\left\langle\left[\delta f_\phi(t)\right]^2\right\rangle&=&
\left\langle\left[\delta f_\phi(0)\right]^2\right\rangle
+\frac{4\varepsilon^2 N^2t^2}{\hbar^2} \left\langle\left[\delta
f_\theta(0)\right]^2\right\rangle,\label{dfp2}\hskip 12pt
\end{eqnarray}
with
\begin{eqnarray}
\left\langle\left[\delta
f_\phi(0)\right]^2\right\rangle=\frac{1-m^2}{8}
\left({1\over \sigma_M^2}+3m^2{1\over \sigma_Y^2}\right).
\end{eqnarray}

If on the other hand, we take a distribution similar
to Eq. (\ref{npj}) but with no correlations between the populations
in the spin components $j=+,0,-$,
\begin{eqnarray}
\varphi(p'_+,p'_0,p'_-)=(2\pi)^{-3/4}(\sigma_+\sigma_0\sigma_-)^{-1/2}\nonumber\\
\times\exp\left[-\frac{1}{4}\left(
\frac{p'^2_+2}{\sigma_+^2}+\frac{p'^2_0}{\sigma_0^2}+\frac{p'^2_-}{\sigma_-^2}
\right)\right],
\end{eqnarray}
with $\sigma_j^2$ denoting the corresponding
atom number variance, we find
\begin{eqnarray}
\left\langle\left[\delta f_r(t)\right]^2\right\rangle&=&
\sigma_+^2+\sigma_0^2+\sigma_-^2,\nonumber\\
\left\langle\left[\delta
f_\theta(t)\right]^2\right\rangle&=&\frac{1}{1-m^2}\left[
(1-m)^2\sigma_+^2+m^2\sigma_0^2\right.\nonumber\\
&&\left.+(1+m)^2\sigma_-^2\right],
\end{eqnarray}
and Eq. (\ref{dfp2}) still holds but with
\begin{eqnarray}
\left\langle\left[\delta
f_\phi(0)\right]^2\right\rangle&=&\frac{1-m^2}{16}
\left[{(1+m)^2\over\sigma_+^2}+4m^2{1\over
\sigma_0^2}\right.\nonumber\\
&&\left.+{(1-m)^2\over\sigma_-^2}\right].
\end{eqnarray}

In the Thomas-Fermi (TF) regime, we obtain most of
the above results analytically.
First we note that the ground wave function is
\begin{eqnarray}
\phi^2(\vec r)=\left\{\begin{array}{cl}
(c_+N)^{-1}[\mu-V_{\rm ext}(\vec r)],&\mbox{if $\mu>V_{\rm ext}(\vec r)$},\\
0,&\mbox{otherwise.}\end{array}\right.,
\end{eqnarray}
where $c_+=4\pi\hbar^2a_2/M$ and the chemical potential
$$\mu=\frac{\hbar\omega_{\rm ho}}{2} \left(\frac{15Na_2}{a_{\rm
ho}}\right)^{2/5}$$ with $\omega_{\rm
ho}=(\omega_x\omega_y\omega_z)^{1/3}$ and $a_{\rm
ho}=(\hbar/M\omega_{\rm ho})^{1/2}$. Similarly,
ignoring the kinetic energy term in the Eq. (\ref{eqphase}),
we obtain
\begin{eqnarray}
\theta(\vec r)=\left\{\begin{array}{cc}
[2V_0\phi(\vec r)]^{-1},&\mbox{if $\mu>V_{\rm ext}(\vec r)$},\\
0,&\mbox{otherwise},
\end{array}\right.
\end{eqnarray}
and
\begin{eqnarray}
\tilde{u}=\frac{c_+}{V_0}
\end{eqnarray}
with $$V_0=\int_{\mu-V_{\rm ext}(\vec r)>0}d\vec r =\frac{4\pi a_{\rm
ho}^3}{3} \left(\frac{15a_2}{a_{\rm ho}}\right)^{3/5}N^{3/5}.$$

All overlap integrals can be calculated analytically as
\begin{eqnarray}
{\cal O}_{\phi\theta}&=&(4V_0)^{-1}=\frac{3}{16\pi a_{\rm ho}^3}
\left(\frac{a_{\rm ho}}{15a_2}\right)^{3/5}N^{-3/5},\nonumber\\
{\cal O}_{\phi\phi}&=&\frac{32\pi\mu^{7/2}}{105c_+^2N^2\omega_{\rm
ho}^3}\left(\frac{2}{M}\right)^{3/2}\nonumber\\
&=&\frac{15^{2/5}}{14\pi a_{\rm ho}^3}\left(\frac{a_{\rm
ho}}{a_2}\right)^{3/5}N^{-3/5}.\nonumber
\end{eqnarray}
The parameter $\varepsilon$ relates essentially to the kinetic
energy operator $E_k^{(\theta)}=-(\hbar^2/2M)\int d\vec
r\theta\nabla^2\theta$ evaluated with respect to the phase function
$\theta(\vec r)$. In the TF approximation we find
\begin{eqnarray}
N\varepsilon=2E_k^{(\theta)}.
\end{eqnarray}
Using Eqs. (\ref{gpe1}) and (\ref{eqphase}) we find
\begin{eqnarray}
({\cal L}-\mu+3c_+N\phi^2)\theta=\frac{N\tilde{u}}{\mu}({\cal
L}+c_+N\phi^2)\phi,
\end{eqnarray}
which simplifies to
\begin{eqnarray}
\left(-\frac{\hbar^2\nabla^2}{2M}+2c_+N\phi^2\right)\theta=\frac{N\tilde{u}}{\mu}
\left(-\frac{\hbar^2\nabla^2}{2M}+\mu\right)\phi,
\label{mp}
\end{eqnarray}
in the TF limit.
Multiplying this Eq. (\ref{mp}) from left by $\theta$ or $\phi$ respectively,
and integrating over $\vec r$, we obtain
\begin{eqnarray}
E_k^{(\theta)}+2c_+N{\cal
O}_{\phi\theta}&=&\frac{N\tilde{u}}{\mu}\left(E_k^{(\phi\theta)}
+\frac{\mu}{2}\right),\nonumber\\
E_k^{(\phi\theta)}+2c_+N\int d\vec
r\phi^3\theta&=&\frac{N\tilde{u}}{\mu}\left(E_k^{(\phi)}
+\mu\right),
\end{eqnarray}
where $E_k^{(\phi)}=-(\hbar^2/2M)\int d\vec r\phi\nabla^2\phi$ is
the kinetic energy per atom in the condensate, and
$E_k^{(\phi\theta)}=-(\hbar^2/2M)\int d\vec
r\phi\nabla^2\theta=-(\hbar^2/2M)\int d\vec r\theta\nabla^2\phi$.
Eliminating $E_k^{(\phi\theta)}$ from above two equations, we then
obtain
\begin{eqnarray}
E_k^{(\theta)}&=&\left(\frac{N\tilde{u}}{\mu}\right)^2E_k^{(\phi)}
+\frac{N\tilde{u}}{2}-2c_+N{\cal O}_{\phi\theta}\nonumber\\
&&+\frac{N\tilde{u}}{\mu}\left(N\tilde{u}-2c_+N\int
d\vec r\phi^3\theta\right)\nonumber\\
&=&\left(\frac{N\tilde{u}}{\mu}\right)^2E_k^{(\phi)}\nonumber\\
&=&\frac{4}{25}E_k^{(\phi)}.\label{ekpt}
\end{eqnarray}
For a spherical trap, adopting the result of Ref. \cite{dalfovo},
we immediately find
\begin{eqnarray}
\varepsilon=\frac{4\hbar^2}{5MNR^2}\ln\left(\frac{R}{1.3a_{\rm
ho}}\right),
\end{eqnarray}
with $$R=(15a_2a_{\rm ho}^4)^{1/5}N^{1/5},$$
which seems to indicate that
$N\varepsilon\propto N^{-2/5}$, asymptotically goes to zero at
large $N$.
By all measures, we find $\varepsilon$ to be a very small quantity
as compared to $c_2{\cal O}_{\phi\theta}$.
Taking $\varepsilon$ as zero in the TF limit, we obtain
\begin{eqnarray}
n_0^2\Delta_{Y0}^2&=&\frac{1}{3}\sqrt{\frac{7}{10}}\,,\nonumber\\
\hbar\Omega&=&-\frac{\sqrt{3/14}}{15^{1/10}}
\left({c_2\over \pi a_{\rm ho}^3}\right)
\left(\frac{a_{\rm ho}}{a_2}\right)^{3/5}N^{2/5}.
\end{eqnarray}

To shed light on the above scaling results based on the
TF approximation, we have performed extensive numerical calculations.
As an example, we consider a spinor condensate of $^{87}$Rb atoms
in a spherically symmetric trap with a harmonic frequency
$\omega=(2\pi)100$ (Hz) and consider the regime of
$N\sim 10^7-10^{15}$. It turns out that our results indeed confirm
the scaling relationship
$${\cal O}_{\phi\phi},{\cal
O}_{\phi\theta},\tilde{u},\Omega/N\propto N^{-3/5},$$ and
\begin{eqnarray}
\frac{N\tilde{u}}{\mu}&\rightarrow& \frac{2}{5},\nonumber\\
n_0^2\Delta_{Y0}^2&\rightarrow&\frac{1}{3}\sqrt{\frac{7}{10}},
\end{eqnarray}
in the large $N$ limit. We also find
that $$E_k^{(\phi)}\propto N^{-0.3549},$$
in rough agreement with the result of Ref. \cite{dalfovo};
On the other hand, we find
$$N\varepsilon\propto N^{0.1348},\quad
E_k^{(\theta)}\propto N^{0.1314},$$
and $$N\varepsilon\approx 4E_k^{(\theta)},
$$ which do not agree with the results above as obtained
with the TF approximation. A careful analysis as confirmed
by our numerical results reveal that although
$\int d\vec r (V_{\rm ext}-\mu+c_+N\phi^2)\phi^2$
is small, $\int d\vec r
(V_{\rm ext}-\mu+c_+N\phi^2)\theta^2$ is not, and approximately
we find
\begin{eqnarray}
E_k^{(\theta)}\approx\frac{1}{2}\int d\vec r (V_{\rm
ext}-\mu+c_+N\phi^2)\theta^2.
\end{eqnarray}
This leads to the opposite $N$-dependence
for $E_k^{(\theta)}$ and $E_k^{(\phi)}$;
while $E_k^{(\theta)}$ increases with increasing values of $N$,
$E_k^{(\phi)}$ decreases. Thus the identity Eq. (\ref{ekpt})
as obtained in the TF limit is invalid.

\section{A nonzero magnetic field}
Inside a nonzero homogeneous magnetic field ($\vec B$),
the atomic quantization axis becomes fixed, along the direction
of $\vec B$, or more conveniently denoted as the z-axis.
Mathematically, it turns out this is equivalent to the
the requirement of the conservation of magnetization
${\cal M}=N_+-N_-$. This reduces the SO(3) rotational symmetry
to its subgroup SO(2), which is still a continuous symmetry.
Thus the quantum dynamics of the phase remains formally
the same as discussed before.

In the linear Zeeman regime, the original Hamiltonian
(\ref{ham}) is augmented to
\begin{eqnarray}
H'=H-\hbar\omega_L{\mathbf F}_z.
\end{eqnarray}
The Larmor precessing
frequency is $\omega_L=B\mu_B/\hbar$ with $\mu_B$ the magnetic
dipole moment for state $|F=1,M_F=1\rangle$.
In this case, the only change is the phases of the spatial
wave functions and the phase functions if of the following
time-dependent form
\begin{eqnarray}
\alpha_j=\alpha_{j0}+j\omega_Lt,
\end{eqnarray}
for $j=+,0,$ and $-$.
$\alpha_{+0}-\alpha_{00}=\alpha_{00}-\alpha_{-0}=0$ still holds
true for a ferromagnetic spinor-1 condensate. A direct consequence
of the magnetic field is thus a time-dependent relative phase
\begin{eqnarray}
\Theta(t)&=&\alpha_{+0}-\alpha_{00}+\omega_Lt\nonumber\\
&=&\alpha_{00}-\alpha_{-0}+\omega_Lt.
\end{eqnarray}
When applied to the macroscopic condensate spin, we find
the spin gains a precession around the z-axis.

Now the picture of the dynamics of the macroscopic condensate spin
is clear: besides the precession induced by the external B-field,
it also diffuses to restore the U(1) and SO(2) symmetries
of the Hamiltonian.

The effect of quadratic Zeeman shift can also be simply addressed.
Apart from an overall shift, it introduces a (positive) B-field
dependent level shift $\propto \hbar\delta_B F_z^2$
to states $|\pm\rangle$ with respect to state $|0\rangle$.
This differential shift causes the precessing of the condensate
spin to be twisted, i.e. the precessing of the $|+\rangle$
($|-\rangle$) component being slower (faster) by $\delta_B$.

\section{The case of anti-ferromagnetic interactions}
Our discussion of the phase diffusion in the previous section
is based on the proof that the single spatial mode approximation
is exact for ferromagnetic interactions.
One might therefore conclude that a similar study for
anti-ferromagnetic interactions would be difficult as,
unlike for ferromagnetic interactions, it's difficult to
predict the ground state condensate structure. In reality, however,
the quantum phase problem for anti-ferromagnetic
interactions is much simpler, provided the mean field description
for the ground state still applies.

First we note that when ${\cal M}\neq 0$, the population in
the $0$-th component is zero in the ground state \cite{pu,su}.
Therefore, the zero mode
Hamiltonian (\ref{hzero2}) reduces to
\begin{eqnarray}
\frac{H_{\rm
zero}}{N}&=&-\frac{P_+^2}{J_+^2}\left(-\frac{J_+}{2}\tilde{u}+2c_-I_{-+}\right)\nonumber\\
&&-\frac{P_-^2}{J_-^2}\left(-\frac{J_-}{2}\tilde{u}+2c_-I_{-+}\right)\nonumber\\
&&+\frac{4c_-P_+P_-}{J_+J_-}\,I_{-+},
\end{eqnarray}
which is exactly the same zero mode Hamiltonian for
a binary condensate with no coupling between its two components.
This has already been studied before \cite{villain97}.
We note that results obtained in Section \ref{model} remain valid.

Now, let's consider the ${\cal M}=0$ case.
As was pointed out in Ref. \cite{su},
the single mode approximation again applies in this case,
and we can assume
\begin{eqnarray}
\phi_j(\vec r)=\sqrt{n_j}\,\phi(\vec r)e^{i\alpha_j},
\end{eqnarray}
where $n_+=n_-=(1-n_0)/2$ and total energy of the system is
independent of the value of $n_0$. A similar relation exists
among the phases of the three components:
$\alpha_++\alpha_--2\alpha_0=\pi$. The spatial profile $\phi(\vec
r)$ satisfies following equation
\begin{eqnarray}
\left[{\cal L}+c_0N\phi^2(\vec r)\right]\phi(\vec r)=\mu\phi(\vec r),
\end{eqnarray}
with a coefficient $c_0$ for the nonlinear term.
The phase function then becomes
\begin{eqnarray}
\theta_j(\vec r)=\sqrt{n_j}\theta(\vec r)e^{i\alpha_j},
\end{eqnarray}
where $\theta(\vec r)$ is the solution of the
following equation
\begin{eqnarray}
\left[{\cal L}-\mu+3c_0N\phi^2(\vec r)\right]\theta(\vec
r)=N\tilde{u}\theta(\vec r).
\end{eqnarray}

Following the results of Sect. \ref{stead}, we define
\begin{eqnarray}
\varepsilon=\frac{2}{N}\int d\vec r\theta(\vec r)\left[{\cal L}
-\mu+c_0N\phi^2(\vec r)\right]\theta(\vec r),\label{epsi}
\end{eqnarray}
and the zero mode Hamiltonian becomes
\begin{eqnarray}
\frac{H_{\rm zero}}{N} &=&\left[\frac{\varepsilon}{(1-n_0){\cal
O_{\phi\theta}}}+2c_0+\frac{2c_2}{1-n_0}\right]
p_+'^2{\cal O}_{\phi\theta}\nonumber\\
&&+\left(\frac{\varepsilon}{2n_0{\cal O_{\phi\theta}}}
+2c_0\right)
p_0'^2{\cal O}_{\phi\theta}\nonumber\\
&&+\left[\frac{\varepsilon}{(1-n_0){\cal
O_{\phi\theta}}}+2c_0+\frac{2c_2}{1-n_0}
\right]p_-'^2{\cal O}_{\phi\theta}\nonumber\\
&&+4c_0p_+'p_0'{\cal O}_{\phi\theta} +4c_0
p_-'p_0'{\cal O}_{\phi\theta}\nonumber\\
&&+4\left(c_0+c_2\frac{2n_0-1}{1-n_0}\right)p_+'p_-'{\cal O}_{\phi\theta}\nonumber\\
&&+\frac{c_2n_0(1-n_0){\cal O}_{\phi\phi}}{2\hbar^2}
\left(q_+'-2q_0'+q_-'\right)^2\nonumber\\
&=&{\mathbf p}'^T{\mathbf{\cal A}{\mathbf p}'} +{\mathbf
q}'^T{\mathbf{\cal B}{\mathbf q}'},\label{hzero3}
\end{eqnarray}
the positive definiteness of matrices ${\mathbf{\cal A}}$ and
${\mathbf{\cal B}}$ can again be verified. By applying the same
transformation (\ref{bt}), we obtain
\begin{eqnarray}
\frac{H_{\rm zero}}{N}&=&ap_N^2+bp_M^2+cp_Y^2+\alpha p_Np_Y+\eta
q_Y^2,\label{antihzero}
\end{eqnarray}
where all coefficients are now given in Appendix \ref{anticoeff}. We
see that fluctuations in $N$ are completely decoupled from
that of $M$ and $Y$ in this case, a result that also
happens for ferromagnetic interactions when ${\cal M}=0$.
The ground state of (\ref{hzero3}) is similar
to Eq. (\ref{hzero2}) except now that
\begin{eqnarray}
\Omega&=&2N\sqrt{c\eta}=\frac{2N}{\hbar}[c_2{\cal
O}_{\phi\phi}(\varepsilon+4c_2n_0^2{\cal
O}_{\phi\theta})]^{1/2},\nonumber\\
\Delta_{Y0}^2&=&\frac{1}{\hbar}\sqrt{\frac{c}{\eta}}=\frac{1}{3}
\left[\frac{\varepsilon+4c_2n_0^2{\cal
O}_{\phi\theta}}{c_2n_0^2(1-n_0)^2{\cal
O}_{\phi\phi}}\right]^{1/2}.
\end{eqnarray}

The dynamic equations for $p_i$ and $q_i$ then become
\begin{eqnarray}
\dot{p}_N&=&0,\nonumber\\
\dot{p}_M&=&0,\nonumber\\
\dot{p}_Y&=&-2N\eta q_Y,\nonumber\\
\dot{q}_N&=&N\left(2ap_N+\alpha p_Y\right),\nonumber\\
\dot{q}_M&=&2Nbp_M,\nonumber\\
\dot{q}_Y&=&N\left(\alpha p_N+2cp_Y\right).
\end{eqnarray}
One can easily find the evolution matrix to be
\begin{widetext}
\begin{eqnarray}
{\mathbf {\cal
T}}(t)=\left(\begin{array}{cccccc}1&0&0&0&0&0\\0&1&0&0&0&0
\\-\alpha'(1-\cos\Omega t)&0&\cos\Omega t
&0&0&-\sin\Omega t/\hbar\Delta_{Y0}^2
\\\zeta_aNt+\hbar\Delta_{Y0}^2\alpha'^2\sin\Omega t&
0& \hbar\Delta_{Y0}^2\alpha'\sin\Omega t&1&
0&-\alpha'(1-\cos\Omega t)
\\0&2bNt&0&0&1&0
\\\hbar\Delta_{Y0}^2\alpha'\sin\Omega t&0
&\hbar\Delta_{Y0}^2\sin\Omega t&0&0&\cos\Omega
t\end{array}\right),\label{antitr}
\end{eqnarray}
\end{widetext}
where
\begin{eqnarray}
\alpha'&=&\frac{\alpha}{2c}=\frac{(3n_0-1)\varepsilon+8c_2n_0^2{\cal
O}_{\phi\theta}}{\sqrt{2}(\varepsilon+4c_2n_0^2{\cal
O}_{\phi\theta})},\nonumber\\
\zeta_a&=&\frac{4ac-\alpha^2}{2c}=12c_0{\cal
O}_{\phi\theta}+\frac{3\varepsilon(\varepsilon+4c_2n_0{\cal
O}_{\phi\theta})}{\varepsilon+4c_2n_0^2{\cal
O}_{\phi\theta}}.\nonumber
\end{eqnarray}
As for $\delta\vec f(t)$ we find
\begin{eqnarray}
\delta f_x&=&\cos\Theta\sqrt{\frac{2n_0}{1-n_0}}p_M
+\sin\Theta\frac{\sqrt{6n_0(1-n_0)}}{\hbar}q_Y,\nonumber\\
\delta f_y&=&-\sin\Theta\sqrt{\frac{2n_0}{1-n_0}}p_M
+\cos\Theta\frac{\sqrt{6n_0(1-n_0)}}{\hbar}q_Y,\nonumber\\
\delta f_z&=&\sqrt{2}p_M.
\end{eqnarray}

\section{conclusions}\label{con}
In conclusion, we have investigated in detail quantum phase
diffusions of a spinor-1 condensate. When the elastic
atomic interaction is of ferromagnetic type, the structure of the
ground state is greatly simplified: the spatial
mode functions of the different spin components
are exactly the same and
accordingly we also proved that the phase functions are exactly the same.
This simplification allows us to construct the zero mode Hamiltonian
which describes the number and phase fluctuations of a condensate
while maintaining the conservations of the
total number of atoms and the total magnetization. We
have provided analytical results for the number and phase fluctuations
due to both the quantum phase diffusion dynamics and the initial
distribution of atom number and phase fluctuations. The structure
of these fluctuations is rather simple: along with the
oscillations due to $\sin\Omega t$ and $\cos\Omega t$ terms,
quantum phase diffusion terms proportional to $Nt$ also exist.
This suggests that the mean-field approach to spinor-1 condensates
is only valid for a finite duration. Based on these results, we
have studied the diffusion of the direction of a macroscopic
condensate spin. Our investigation sheds important light
on the studies of beyond mean field theory quantum
correlations among Bose condensed atoms.

\section{Acknowledgements}
We thank Yueheng Lan and Dr. X. X. Yi for helpful discussions.
This work is supported by NSF
and by a grant from the National Security Agency (NSA),
Advanced Research and Development Activity (ARDA), and the Defense
Advanced Research Projects Agency (DARPA) under Army Research Office
(ARO) Contract No. DAAD19-01-1-0667.

\appendix
\section{overlap integrals}\label{over}
In this appendix, we list the overlap integrals
as involved in the dynamic Eq. (\ref{dynpq}).
Using the phase conventions as introduced in
Eqs. (\ref{wave}) and (\ref{phase}),
we find explicitly
\begin{eqnarray}
J_j&=&n_j,\nonumber\\
I_{jk}&=&n_jn_k{\cal O}_{\phi\theta},\nonumber\\
(\phi_0\theta_0|\phi_-\theta_+)
&=&(\phi_0\theta_0|\phi_+\theta_-)\nonumber\\
&=&(\phi_0^2|\theta_+\theta_-)\nonumber\\
&=&n_0^2{\cal O}_{\phi\theta}/2\nonumber\\
&=&2n_+n_-{\cal O}_{\phi\theta},\nonumber\\
(\phi_0^2|\phi_+\phi_-)&=&n_0^2{\cal O}_{\phi\phi}/2.
\end{eqnarray}

\section{The zero mode Hamiltonian}\label{zero}
The zero mode Hamiltonian is listed below.
\begin{widetext}
\begin{eqnarray}
\frac{H_{\rm
zero}}{N}&=&-\frac{P_+^2}{J_+^2}\left[-\frac{J_+}{2}\tilde{u}+2c_-I_{-+}+2c_+I_{+0}
+2c_2(\phi_0\theta_0|\phi_-\theta_+)'
+c_2(\phi_0^2|\theta_+\theta_-)'\right]\nonumber\\
&&-\frac{2P_0^2}{J_0^2}\left[-\frac{J_0}{4}\tilde{u}+c_+I_{+0}+c_+I_{-0}
+c_2(\phi_-\theta_+|\phi_0\theta_0)'
+c_2(\phi_+\theta_-|\phi_0\theta_0)'\right]\nonumber\\
&&-\frac{P_-^2}{J_-^2}\left[-\frac{J_-}{2}\tilde{u}+2c_-I_{-+}+2c_+I_{-0}+2c_2(\phi_0\theta_0|\phi_+\theta_-)'
+c_2(\phi_0^2|\theta_+\theta_-)'\right]\nonumber\\
&&+\frac{4P_+P_0}{J_+J_0}\left[c_+I_{+0}
+c_2(\phi_0\theta_0|\phi_-\theta_+)'\right]+\frac{4P_-P_0}{J_-J_0}\left[c_+I_{-0}
+c_2(\phi_0\theta_0|\phi_+\theta_-)'\right]\nonumber\\
&&+\frac{2P_+P_-}{J_+J_-}\left[2c_-I_{-+}
+c_2(\phi_0^2|\theta_+\theta_-)'\right]
-\frac{c_2}{\hbar^2}(\phi_0^2|\phi_+\phi_-)'
\left(\frac{Q_+}{J_+}-\frac{2Q_0}{J_0} +\frac{Q_-}{J_-}\right)^2.
\end{eqnarray}
\end{widetext}

\section{Coefficients in the zero mode Hamiltonian for ferromagnetic interactions}\label{coeff}
The coefficients in the zero mode Hamiltonian (\ref{hzero}) are
\begin{eqnarray}
a&=&\frac{1}{4n_0^2}\bigglb[\frac{\varepsilon}{3}(5+3m^2)\nonumber\\
&&+2{\cal O}_{\phi\theta}\left(3c_0(1-m^2)^2
+\frac{8}{3}c_2(1-3m^2)\right)\biggrb],\nonumber\\
b&=&\frac{1}{2n_0^2}\left[\varepsilon(1+m^2)
-8c_2{\cal O}_{\phi\theta}m^2\right]\nonumber\\
c&=&\frac{1}{3n_0^2}(\varepsilon-4c_2{\cal O}_{\phi\theta}),\nonumber\\
\alpha&=&\frac{\sqrt{2}}{6n_0^2}(1+3m^2)
(\varepsilon-4c_2{\cal O}_{\phi\theta}),\nonumber\\
\beta&=&-\frac{2m}{\sqrt{3}n_0^2}(\varepsilon
-4c_2{\cal O}_{\phi\theta}),\nonumber\\
\gamma&=&-\frac{2m}{n_0^2}\sqrt{\frac{2}{3}}\left[
\varepsilon-c_2{\cal O}_{\phi\theta}(1+3m^2)\right],\nonumber\\
\eta&=&-3c_2n_0^2{\cal O}_{\phi\phi}/\hbar^2.
\end{eqnarray}

\section{Parameters in Evolution matrix}\label{tran}
The list of parameters in the evolution matrix ${\mathbf{\cal
T}}(t)$ (\ref{tr}) are
\begin{eqnarray}
\alpha'&=&\frac{\alpha}{2c}=\frac{1+3m^2}{2\sqrt{2}},\nonumber\\
\beta'&=&\frac{\beta}{2c}=-\sqrt{3}m,\nonumber\\
\zeta_a&=&\frac{4ac-\alpha^2}{2c}=\frac{3(1+m^2)\varepsilon}{1-m^2}
+12(c_0+c_2){\cal O}_{\phi\theta},\nonumber\\
\zeta_b&=&\frac{4bc-\beta^2}{2c}=\frac{4\varepsilon}{1-m^2},\nonumber\\
\zeta_c&=&\frac{2c\gamma-\alpha\beta}{2c}
=-\frac{2\sqrt{6}m\varepsilon}{1-m^2}.
\end{eqnarray}

\section{Explicit form of number and phase fluctuation}\label{time}
Here we present the explicit form of the solutions for the
number and phase fluctuations.
\begin{widetext}
\begin{eqnarray}
p_N(t)&=&p_N(0),\nonumber\\
p_M(t)&=&p_M(0),\nonumber\\
p_Y(t)&=&-\frac{1+3m^2}{2\sqrt{2}}(1-\cos\Omega t)p_N(0)
+\sqrt{3}m(1-\cos\Omega t)p_M(0)+p_Y(0)\cos\Omega t
-\frac{q_Y(0)}{\hbar\Delta_{Y0}^2}\sin\Omega t,\nonumber\\
q_N(t)&=&\left[Nt\left(\frac{3(1+m^2)\varepsilon}{1-m^2}
+12(c_0+c_2){\cal O}_{\phi\theta}\right)+\frac{(1+3m^2)^2}{8}
\hbar\Delta_{Y0}^2\sin\Omega t\right]p_N(0)\nonumber\\
&&-\left[Nt\frac{2\sqrt{6}m\varepsilon}{1-m^2}+\frac{\sqrt{3}m(1+3m^2)}
{2\sqrt{2}}\hbar\Delta_{Y0}^2\sin\Omega t\right]p_M(0)\nonumber\\
&&+\frac{1+3m^2}{2\sqrt{2}}\hbar^2\Delta_{Y0}^2p_Y(0)\sin\Omega t
+q_N(0)-\frac{1+3m^2}{2\sqrt{2}}(1-\cos\Omega t)q_Y(0),\nonumber\\
q_M(t)&=&-\left[Nt\frac{2\sqrt{6}m\varepsilon}{1-m^2}+\hbar\Delta_{Y0}^2
\frac{\sqrt{3}m(1+3m^2)}{2\sqrt{2}}\sin\Omega t\right]p_N(0)\nonumber\\
&&+\left(Nt\frac{4\varepsilon}{1-m^2}+3m^2\hbar\Delta_{Y0}^2\sin\Omega
t \right)p_M(0)-\sqrt{3}m\hbar\Delta_{Y0}^2p_Y(0)\sin\Omega t
+q_M(0)+\sqrt{3}m(1-\cos\Omega t)q_Y(0),\nonumber\\
q_Y(t)&=&\hbar\Delta_{Y0}^2\frac{1+3m^2}{2\sqrt{2}}p_N(0)\sin\Omega
t -\hbar\Delta_{Y0}^2\sqrt{3}mp_M(0)\sin\Omega t
+\hbar\Delta_{Y0}^2p_Y(0)\sin\Omega t+q_Y(0)\cos\Omega
t,\nonumber\\
p_+'(t)&=&\frac{3(1-m^2)+(1+3m^2)\cos\Omega
t}{4\sqrt{3}}p_N(0) +\frac{1+m-m\cos\Omega t}{\sqrt{2}}p_M(0)
+\frac{\cos\Omega t}{\sqrt{6}}p_Y(0)
-\frac{\sin\Omega t}{\sqrt{6}\hbar\Delta_{Y0}^2}q_Y(0),\nonumber\\
p_0'(t)&=&\frac{3(1+m^2)-(1+3m^2)\cos\Omega t}{2\sqrt{3}}p_N(0)
-\sqrt{2}m(1-\cos\Omega t)p_M(0)
-\frac{2\cos\Omega t}{\sqrt{6}}p_Y(0)
+\frac{2\sin\Omega t}{\sqrt{6}\hbar\Delta_{Y0}^2}q_Y(0),\nonumber\\
p_-'(t)&=&\frac{3(1-m^2)+(1+3m^2)\cos\Omega t}{4\sqrt{3}}p_N(0)
-\frac{1-m+m\cos\Omega t}{\sqrt{2}}p_M(0) +\frac{\cos\Omega
t}{\sqrt{6}}p_Y(0)
-\frac{\sin\Omega t}{\sqrt{6}\hbar\Delta_{Y0}^2}q_Y(0),\nonumber\\
q_+'(t)&=&\left[\sqrt{3}Nt\left(\frac{(1-m)\varepsilon}{1+m}+4(c_0+c_2)
{\cal O}_{\phi\theta}\right)+\frac{\sqrt{3}\hbar\Delta_{Y0}^2}{8}
(1+3m^2)(1+m)^2\sin\Omega t\right]p_N(0)\nonumber\\
&&+\left[\frac{2\sqrt{2}\varepsilon}{1+m}Nt
-\frac{3m(1-m)^2}{2\sqrt{2}}\hbar\Delta_{Y0}^2\sin\Omega
t\right]p_M(0)
+\hbar\Delta_{Y0}^2\frac{3(1-m)^2\sin\Omega t}{2\sqrt{6}}p_Y(0)\nonumber\\
&&+\frac{1}{\sqrt{3}}q_N(0)+\frac{1}{\sqrt{2}}q_M(0)
+\left[-\frac{1-6m+3m^2}{2\sqrt{6}}
+\frac{(1-m)^2}{2\sqrt{2}}\cos\Omega t\right]q_Y(0),\nonumber\\
q_0'(t)&=&\left[\sqrt{3}Nt\left(\frac{1+m^2}{1-m^2}\varepsilon
+4(c_0+c_2){\cal O}_{\phi\theta}\right)+\frac{\sqrt{3}}{8}
\hbar\Delta_{Y0}^2(3m^4-2m^2-1)\sin\Omega t\right]p_N(0)\nonumber\\
&&+\left[-\frac{2\sqrt{2}m\varepsilon}{1-m^2}Nt
+\hbar\Delta_{Y0}^2\frac{3m(1-m^2)}{2\sqrt{2}}\sin\Omega
t\right]p_M(0)
+\hbar\Delta_{Y0}^2\frac{3m^2-3}{2\sqrt{6}}\sin\Omega p_Y(0)\nonumber\\
&&+\frac{1}{\sqrt{3}}q_N(0)+\left[-\frac{1+3m^2}{2\sqrt{6}}
+\frac{3m^2-3}{2\sqrt{6}}\cos\Omega t\right]q_Y(0),\nonumber\\
q_-'(t)&=&\left[\sqrt{3}Nt\left(\frac{(1+m)\varepsilon}{1-m}+4(c_0+c_2)
{\cal O}_{\phi\theta}\right)+\frac{\sqrt{3}\hbar\Delta_{Y0}^2}{8}
(1+3m^2)(1+m^2)\sin\Omega t\right]p_N(0)\nonumber\\
&&-\left[\frac{2\sqrt{2}\varepsilon}{1-m}Nt+\frac{3m(1+m)^2}{2\sqrt{2}}
\hbar\Delta_{Y0}^2\sin\Omega t\right]p_M(0)
+\hbar\Delta_{Y0}^2\frac{3(1+m)^2\sin\Omega t}{2\sqrt{6}}p_Y(0)\nonumber\\
&&+\frac{1}{\sqrt{3}}q_N(0)-\frac{1}{\sqrt{2}}q_M(0)
+\left[-\frac{1+6m+3m^2}{2\sqrt{6}}
+\frac{(1+m)^2}{2\sqrt{2}}\cos\Omega t\right]q_Y(0).
\end{eqnarray}
\end{widetext}

\section{The variance of macroscopic condensate spin in Cartesian
coordinates} \label{car}
In Cartesian coordinates,
the variances of the condensate spin direction
is given by
\begin{widetext}
\begin{eqnarray}
\left\langle\left[\delta
f_x(t)\right]^2\right\rangle&=&\frac{1}{1-m^2}\left[3\langle
p_N^2(0)\rangle-2\sqrt{6}m\langle p_N(0)p_M(0)\rangle
+2m^2\langle p_M^2(0)\rangle\right],\nonumber\\
%%%%%%%%%%%%%%%%%%%%%%%%%%%%%%%%%%%
\left\langle\left[\delta
f_y(t)\right]^2\right\rangle&=&\frac{1-m^2}{2\hbar^2}[\langle
q_M^2(0)+2\sqrt{3}m\langle q_M(0)q_Y(0)\rangle+3m^2\langle
q_Y^2(0)\rangle]\nonumber\\
&&+\frac{4N^2\varepsilon^2t^2}{\hbar^2(1-m^2)}[3m^2\langle
p_N^2(0)\rangle-2\sqrt{6}m\langle p_N(0)p_M(0)\rangle
+2\langle p_M^2(0)\rangle]\nonumber\\
&&+\frac{\sqrt{2}N\varepsilon
t}{\hbar^2}\times\left[\sqrt{2}\left(\langle
q_M(0)p_M(0)\rangle+\langle
p_M(0)q_M(0)\rangle\right)-\sqrt{3}m\left(\langle
q_M(0)p_N(0)\rangle
+\langle p_N(0)q_M(0)\rangle\right)\right.\nonumber\\
&&\left.+\sqrt{6}m\left(\langle q_Y(0)p_N(0)\rangle +\langle
p_N(0)q_Y(0)\rangle\right)-3m^2\left(\langle
q_Y(0)p_M(0)\rangle+\langle p_M(0)q_Y(0)\rangle\right)
\right],\nonumber\\
%%%%%%%%%%%%%%%%%%%%%%%%%%%%%%%%%%%
\left\langle\left[\delta f_z(t)\right]^2\right\rangle&=&2\langle
p_M^2(0)\rangle.
\end{eqnarray}
\end{widetext}

\section{Coefficients in the zero mode Hamiltonian for anti-ferromagnetic interactions}\label{anticoeff}
The coefficients of zero mode Hamiltonian (\ref{antihzero})
for anti-ferromagnetic interactions (with ${\cal M}=0$)
\begin{eqnarray}
a&=&\frac{(1+3n_0)\varepsilon}{6n_0(1-n_0)}+\left[6c_0-\frac{8c_2n_0}{3(1-n_0)}\right]
{\cal O}_{\phi\theta},\nonumber\\
b&=&\frac{\varepsilon}{1-n_0}+4c_2{\cal O}_{\phi\theta},\nonumber\\
c&=&\frac{\varepsilon}{3n_0(1-n_0)}+\frac{4c_2n_0}{3(1-n_0)}{\cal O}_{\phi\theta},\nonumber\\
\alpha&=&\frac{2(3n_0-1)\varepsilon}{3\sqrt{2}n_0(1-n_0)}
+\frac{16c_2n_0}{3\sqrt{2}(1-n_0)}{\cal O}_{\phi\theta},\nonumber\\
\eta&=&3c_2n_0(1-n_0){\cal O}_{\phi\phi}/\hbar^2.
\end{eqnarray}

\end{document}